# Water megamasers and the central black hole masses in a large sample of galaxies

Ahlam Farhan [a,\*], Enise Nihal Ercan [b], Francesco Tombesi [c,d,e,f,g]

[a] *Currently Self-Employed, Turkeli Street, Pendik, Istanbul 34903, Turkey*
[b] *Department of Physics, Bogazici University, Istanbul 34342, Turkey*
[c] *Department of Physics, Tor Vergata University of Rome, Via della Ricerca Scientifica 1, I-00133 Rome, Italy*
[d] *INAF - Astronomical Observatory of Rome, Via Frascati 33, I-00078 Monte Porzio Catone, Rome, Italy*
[e] *INFN - Roma Tor Vergata, Via della Ricerca Scientifica 1, I-00133 Rome, Italy*
[f] *Department of Astronomy, University of Maryland, College Park, MD 20742, USA*
[g] *NASA/Goddard Space Flight Center, X-ray Astrophysics Laboratory, Greenbelt, MD20771, USA*

**Abstract**

Extragalactic water maser emissions at 22 GHz have been playing vital roles in astrophysics. The limited detection rate of these masers has been motivating researchers to find clues that can help characterise them. The physical environments 22 GHz masers formed in are still ambiguous. Accordingly, statistical studies have been thoroughly used to resolve these favourable environments. This work goes through the essential parameter of Active Galactic Nuclei (AGN), namely, the mass of the central supermassive black hole ($M_{BH}$) of the maser host galaxy. We study the correlation between maser luminosity ($L_{H_2O}$) and $M_{BH}$ in sub-samples of megamasers (MMs), kilomasers (KMs), and disc masers. The regression line of the relation is also calculated for these sub-samples. Unlike the results of previous works, dividing the maser sample into MMs and KMs gives no privilege to MM galaxies. Contrary to expectation, KMs have weak and low significant $L_{H_2O}$ - $M_{BH}$ correlation, while MMs show no correlation. The positive correlation in KMs can be explained by the role of AGN therein, while the diversity of MMs types, with some of which are not strongly related to AGN, may explain the correlation missing. Surprisingly, the 28 disc maser sample, where tight correlation is expected, shows a very weak and low significant $L_{H_2O}$-$M_{BH}$ correlation. Future VLBI studies will eventually lead to a specific classification of a good number of maser galaxies, which is essential to establishing the $L_{H_2O}$-$M_{BH}$ relation.

## 1. Introduction

Since their discovery at the centre of NGC 4945 galaxy in 1979 (Santos and Lépine, 1979), the microwave amplification by stimulated emission of radiation (maser) emitted by water molecules at 22 GHz from external galaxies has provided insights into several fundamental problems in astrophysics. The unique environments in which these water masers are produced have allowed the most accurate way to calculate black hole masses outside our galaxy (e.g., Miyoshi et al., 1995; Greenhill and Gwinn, 1997; Herrnstein et al., 1999; Kondratko et al., 2005; Reid et al., 2009; Kuo et al., 2010; Gao et al., 2016; Kuo et al., 2020). More exceptionally, the sub-milliarcsecond resolution of radio telescopes (e.g., the Very Long Baseline Array (VLBA); the Atacama Large Millimeter/submillime-

⇑ Corresponding author.
*E-mail addresses:* ahlamfarhan@yahoo.com (A. Farhan), ercan@boun.edu.tr (E.N. Ercan), francesco.tombesi@roma2.infn.it (F. Tombesi).

ter Array (ALMA)) has made it possible to track water maser spots within the circumnuclear disc (CND) vicinity, just a few parsecs from the central engine of galaxies hosting Active Galactic Nuclei (AGN), which have been used since 1999 (e.g., Herrnstein et al., 1999) in distance calculations (e.g., Kuo et al., 2010; Gao et al., 2016). Calculating the distance to the water maser host galaxies independently in one-step, without referring to the standard candles or using the cosmic microwave background, has provided a 4% constraint on the Hubble parameter ($H_o$) (see Pesce et al., 2020 and references therein).

The Megamaser Cosmology Project (MCP[1]; summary of this project can be found in Braatz et al., 2017) is a multi-year project run by the National Radio Astronomy Observatory (NRAO, 1956) to measure $H_o$ with an accuracy of 3%. The team of MCP has been finding, monitoring, and mapping water maser spots at centres of AGN within the accretion discs therein. These observations allow direct distance measurement of each galaxy with about 10% accuracy for each studied galaxy. Together, ten galaxies with such an accuracy will help the MCP team to reach the wanted 3% constraint on $H_o$. Calculating $H_o$ with this minimal accuracy will help constrain the equation of the state of dark energy, which is a vital issue in modern cosmology. The MCP has reached a 4% accuracy to $H_o$ (Pesce et al., 2020), and the team of MCP is now very close to the initially assigned goal. However, increasing the detection rate of disc water maser galaxies is needed to accelerate the MCP's results.

Unfortunately, currently, the detection rate of water masers is very small. Up to now, more than 6000 galaxies have been searched for 22 GHz water maser with roughly 200 discovered. Researchers are keen to find hints that may help distinguish maser host galaxies. Among such successful suggestions is the work of Henkel et al. (2005), where they used high far-infrared luminosity and central-based jet galaxies as selection criteria to detect masers. These two characteristics alone allowed a detection rate of 50%. Another good criterion is high nuclear X-ray obscuring column density ($n_H$), galaxies with $n_H > 10^{22}$ cm$^{-2}$ were found by Huré et al. (2011) to be more likely maser-hosts. Other works have also shown tentative results, such as the one that chose galaxies with high radio (20 cm) luminosity (Liu et al., 2016), or luminous X-ray galaxies (Wagner, 2013).

Extragalactic water masers are usually classified as Kilomasers (KMs) and Megamasers (MMs). Galaxies with maser luminosity $L_{H_2O} < 10\ L_\odot$ are KMs, whereas those with $L_{H_2O} \geqslant 10\ L_\odot$ are MMs (Tarchi et al., 2010). KMs are thought be powered by star formation processes, where a superposition of several stellar maser spots is adequate to explain the total maser output flux, and this interpretation is supported by the fact that KMs are observed in intense star forming regions, such as those found in the Antennae interacting galaxies NGC 4038/NGC 4039 (Brogan et al., 2010). On the contrary, the immense flux of MMs can not be produced from stellar sources. The exclusive occurrence of MMs at centres of galaxies has led to a consensus that they are powered by AGN (e.g., Lo, 2005; Constantin, 2012). A subclass of MMs are disc masers, and they have 3-featured spectra, i.e., redshifted, systematic, and blue-shifted spectra (Kuo et al., 2010). Disc masers emission is thought to be produced from a CND rotating in a Keplerian orbit at radii of 0.2–1 pc around the central black hole (Greene et al., 2016) thus; it is helpful to study disc masers as a sub-sample to compare the effect of AGN on the maser emission therein. Of the discovered water maser galaxies, about two-thirds are MMs, and at least 30 galaxies of MMs are disc masers (Braatz, 2018).

Since water masers, particularly MMs, are believed to be powered by the central engine of AGN, studying AGN-related characteristics of the host galaxies may also reveal interesting results and thus lead to the long-awaited increase in the detection rate. In this work, we try to re-tackle the essential parameter of an AGN, which is the mass of the central supermassive black hole, $M_{BH}$, and how it correlates with water maser luminosity, $L_{H_2O}$. The paper is ordered as follows: in Section 2 we review previous works and mark out our motivations. Section 3 describes the dataset. Methods used in the dataset analysis are briefly mentioned in Section 4. We demonstrate our analysis results in Section 5. Our conclusion is drawn in Section 7.

## 2. Motivations and previous works

Two published studies systematically examined the $L_{H_2O}$-$M_{BH}$ relationship for a statistically good sample number. The first was in Su et al. (2008), with a sample of 38 maser galaxies with their maser and hard (2–10 keV) X-ray luminosities. This work found a strong $L_{H_2O}$-$M_{BH}$ correlation for a 34 galaxies sample. Specifically, their regression analysis showed a $L_{H_2O} \propto M_{BH}^{3.6}$ line fit for the relation. Although this proportionality can, to some extent, be explained by the theoretically expected quadratic result, $L_{H_2O} \propto M_{BH}^2$, the method used to calculate it was mathematically incorrect (see comment on this in Kandalyan and Al-Zyout, 2010). In that paper, it was impossible to examine disc maser galaxies with just seven available disc galaxies' central stellar velocity dispersion, $\sigma$, data. The MMs sub-sample was not studied.

In the second study, Kandalyan and Al-Zyout (2010) restudied this relation for 39 galaxies and confirmed the strong correlation. They divided their sample into KMs and MMs. KMs (just eight galaxies) showed no correlation, while MMs sub-sample (31 galaxies) displayed a moderate correlation. Spearman's rank correlation coefficient ($\rho$) calculated was $\rho = 0.52$, and the p-value was p = 2x10$^{-3}$. The regression calculated for the relation did not show the quadratic relation. Instead, the proportionality was $L_{H_2O} \propto M_{BH}^{0.6}$. However, in that paper, $M_{BH}$ values were collected from the literature, which means that they were affected by the variety of methods used, increasing the uncertainty level.

---

[1] https://safe.nrao.edu/wiki/bin/view/Main/MegamaserCosmologyProject.

Considering the above-noted results, the correlation between maser and $M_{BH}$ might be advantageous in estimating black hole masses, especially for Type 2 AGN, where a high obscuring level from CND makes $M_{BH}$ calculation impossible using other methods. Besides, there may be potential clues from this relation that will help in future maser surveys; we try to re-investigate this correlation with a sample with an increase of almost double. Additionally, a sample of 28 disc galaxies is separately studied here for the first time.

## 3. The sample and selection criteria

We cross-matched all maser galaxies with observed $L_{H_2O}$ from Braatz (2018) and Kuo et al. (2018) with central velocity dispersion ($\sigma$) data from the Hyper-LEDA database (Hyper-LEDA) and then checked extra available $\sigma$ in the literature. A total sample of 73 maser galaxies was found. Then, we calculated the black hole masses of the host galaxy using the well-known $M_{BH}$–$\sigma$ relation:

$$\log(M_{BH}/M_\odot) = \alpha + \beta \log(\sigma/\sigma_o) \quad (1)$$

where $M_\odot$ is solar mass, the values of 8.13, 4.02, and 200 kms$^{-1}$ for $\alpha$, $\beta$, and $\sigma_o$, respectively, were adopted as estimated in Tremaine et al. (2002).

Table A.1 (see Appendix A) reports the 73 galaxies sample. The table includes types of masers, redshift (z), maser isotropic luminosity $L_{H_2O}$ the integrated flux of all maser spectra in a source is adopted. The details of the calculation are described in Pesce et al. (2015), $\sigma$, calculated $M_{BH}$, and references for the data.

Theoretically, since disc masers emission is thought to be produced from a CND, and MMs are thought to be powered directly by AGN, as explained in Section 1, it is helpful to study MMs and their disc masers sub-sample to compare the effect of AGN on the maser emission therein.

In our analysis, disc masers are either masers with three-featured or double-featured spectra sources (i.e., galaxies with blueshifted redshifted but no systematic spectra) adopted from Kuo et al. (2020). We include the sub-samples of MMs (56 galaxies), KMs (17 galaxies), and disc masers (28 galaxies) in our calculation.

The histogram distributions of $L_{H_2O}$ and $M_{BH}$ for the whole sample are shown in Fig. 1. The mean values of $\log(L_{H_2O}/L_\odot)$ and $\log(M_{BH}/M_\odot)$ are 1.47 and 7.38, respectively. The maximum value of $\log(L_{H_2O}/L_\odot)$ is 3.36 for the merger AGN galaxy NGC 5765b and the minimum is −2 in the starburst galaxy IC 342. The most massive black hole in the sample resides at the centre of NGC 2639 galaxy, with a mass of the order $10^9$ $M_\odot$, the dwarf barred irregular galaxy NGC 4214 has the lightest $M_{BH}$ of 5.6X$10^5$ $M_\odot$.

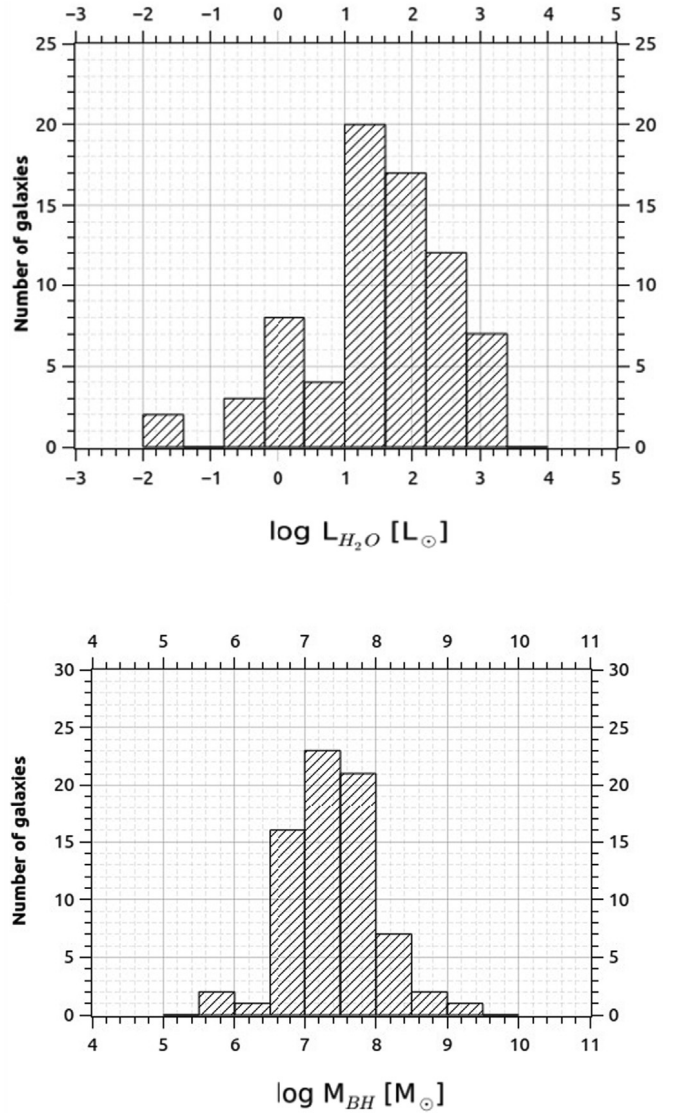

Fig. 1. Histogram distribution of $L_{H_2O}$ (upper panel) and $M_{BH}$ (lower panel) data on a logarithmic scale.

## 4. Analysis of the data

Two approaches to calculate how variables correlate with each other exist: parametric and non-parametric. In parametric analysis, one presupposes that the data in hand has a normal distribution. The Pearson method is the most common parametric method, where the degree of correlation is measured by Pearson correlation coefficient (r). On the other hand, non-parametric approaches do not require any presumptions, which is the main reason behind their wide usage in astrophysics, where specific physical models barely exist to the degree that we can claim normality. We should also mention that non-parametric methods can minimise the effect of the outliers since they measure the correlation of the variables' ranks after ordering them instead of their values. The Spearman's rank correlation coefficient, $\rho$, is the most commonly used non-parametric correlation coefficient.

In our analysis, we first tested the normality of the data in Table A.1 using the Kolmogorov–Smirnov test, and we got p-values of 0.41 and 0.87 for Log($L_{H_2O}$) and Log($M_{BH}$), respectively, higher than the threshold p-value of 0.05 for statistical significance, which means that our data set does not differ significantly from the normal distribution. Thus, we use the parametric Pearson correlation method to analyse the whole sample. The tricky point - unfortunately not always met in published papers- is the so-called Malmquist effect, or the fake correlation one may get between two variables when each depends on redshift (z). This effect is typical when studying luminosities (e.g., Butkevich et al., 2005). Fortunately, using partial correlation coefficients allows us to eliminate this effect by studying the correlation of variables while keeping z constant.

We use the "ppcor" package of the Central R Archive Network (R CRAN; Kim, 2015), to calculate Spearman and Pearson partial correlation coefficients, hereafter $\rho_{partial}$, and $r_s$, respectively. These partial coefficients are calculated when both variables depend on z. In this paper, wherever r is calculated, it means that the sample has a normal distribution, and the size of it is enough (n > 30), so the effect of the outliers can be neglected (Kandalyan and Al-Zyout, 2010). Otherwise, $\rho$ is a better choice. All logarithms are to the base 10, $M_{BH}$ is in solar mass units ($M_\odot$), and isotropic maser luminosities are in solar luminosity units ($L_\odot$).

In addition to correlations, the regression line of the log($L_{H_2O}$)-log($M_{BH}$) relation is calculated to check compatibility with theory, where the quadratic relation expects a slope of 2 for the line (e.g., Neufeld et al., 1994; Su et al., 2008).

## 5. Results of the data analysis

MMs are thought to be directly powered by AGN (Lo, 2005). Thus the correlation is expected to be tight in these galaxies. However, we found no log($L_{H_2O}$)-log($M_{BH}$) correlation in MMs (p = 0.17, p = 0.202); even if we consider Spearman's rank coefficient, $\rho$, in case outliers exist so their effect is minimised, still, no correlation is found ($\rho$ = 0.253, p = 0.060), the scatter plot for the MMs sample is shown in Fig. 2. In contrast, for KMs, a moderate but low significant correlation is found ($\rho$ = 0.60, p = 0.012). The regression line of log($L_{H_2O}$)-log($M_{BH}$) for the KMs sample has the following best fit,

$$log(L_{H_2O}) = 0.83 log(M_{BH}) - 5.77 \quad (2)$$

the standard error (S) of the intercept is 2.03, and the 68% confidence interval has a lower limit of −7.87 and an upper limit of −3.69, while the slope has a standard error of 0.29, lower limit of 0.52, and upper limit of 1.31.

Disc galaxies sub-sample (28 galaxies) has a very weak and low significant ($L_{H_2O}$)-log($M_{BH}$) correlation, with $\rho$ = 0.41 and p = 0.03. The regression line of the disc sample follows the equation,

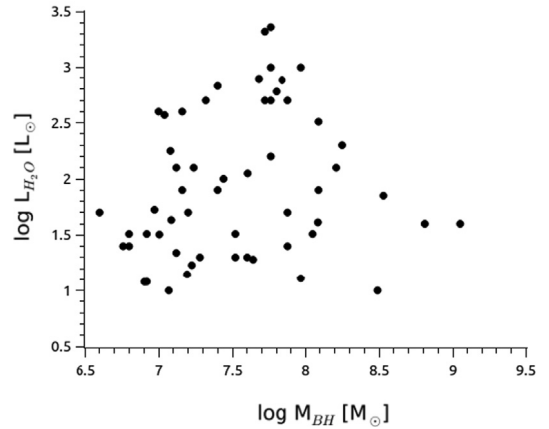

Fig. 2. Scatter plot of the isotropic maser luminosity and the mass of the central black hole of the host galaxy for the 56 MM galaxies..

$$log(L_{H_2O}) = 0.73 log(M_{BH}) - 3.24 \quad (3)$$

the intercept has S = 2.25, and the 68% confidence interval has a lower limit of −5.56 and an upper limit of −1.00, while the slope has a standard error of 0.30, a lower limit of 0.44, and an upper limit of 1.04. Eq. (2) and (3) are displayed in Fig. 3 and 4, respectively. These two equations differ from the theoretically expected quadratic equation.

## 6. Discussion

The AGN power effect on maser emission was expected theoretically by Neufeld et al. (1994). However, the link between AGN parameters (e.g., [O III] $\lambda$ = 5007Å, and X-ray luminosities) and maser emission has not yet been explicitly proved (e.g., Leiter et al., 2017; Zhu et al., 2011). Instead, general trends have been found between maser detection rate and higher X-ray luminosities, larger $\sigma$, and higher [O III] luminosities.

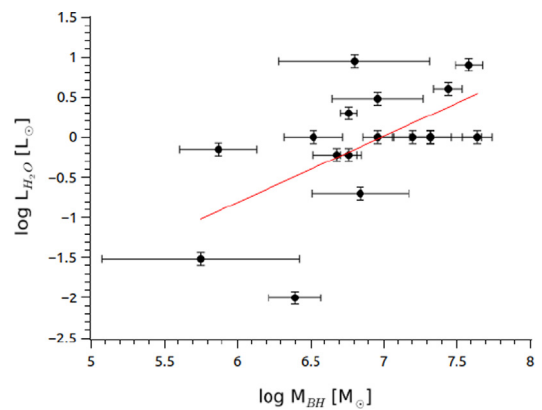

Fig. 3. Relation between the isotropic maser luminosity and the mass of the central black hole of the host galaxy for the 17 KM galaxies. The red line is the linear fit. Since no uniform errors are provided for maser luminosities in the literature, a 20% uncertainty is adopted accounting for maser flux calibration errors (e.g., see Panessa et al., 2020). Error bars in $M_{BH}$ come from uncertainty in $\sigma$.

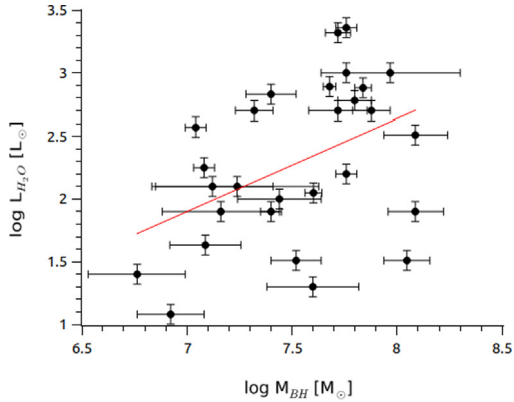

Fig. 4. Relation between the isotropic maser luminosity and the mass of the central black hole of the host galaxy for the 28 disc galaxies. The red line is the linear fit. Error bars are calculated same as Fig. 3. (For interpretation of the references to colour in this figure legend, the reader is referred to the web version of this article.)

Theoretically, water vapour molecules irradiated by energetic X-ray photons from material accreting onto a central supermassive black hole (SMBH) are efficiently excited to emit maser (Neufeld et al., 1994). The interrelation between $L_{H_2O}$ and $M_{BH}$ can be understood from the following picture: nuclear masing actions take place within a sphere of a critical radius $R_{crit}$, within which CND becomes molecular, whereas clouds of higher radius are atomic due to high ionisation levels, thus not appropriate for masing. $L_{H_2O}$ has a direct proportionality with $R_{crit}$ (as proposed by Neufeld et al., 1994), and $R_{crit}$, in turn, depends on 2–10 keV X-ray luminosity ($L_{2-10}$), $\dot{M}$, and $M_{BH}$ (as in Eq. 4 of Neufeld and Maloney, 1995). If we assume that $L_{2-10}$ is proportional to AGN's luminosity ($L_{AGN}$), and also $\dot{M}$ is proportional to $L_{AGN}$ (e.g., Frank et al., 2002), then $R_{crit} \propto (L_{AGN}).(M_{BH})$. Finally, considering the mass-to energy conversion energy (e.g., $L_{AGN} \propto \dot{M}$), we conclude the direct dependency of $L_{H_2O}$ on $M_{BH}$. This correlation is expected in MMs, especially those of disc masers, where maser spots are located at 0.1–1 pc of the central SMBH and directly pumped by AGN's X-ray.

Based on a sample up to double the size of previous works (Kandalyan and Al-Zyout, 2010; Su et al., 2008), we found that maser emissions and central black hole masses, as calculated from σ, do not correlate in MMs, and their correlation is very weak in disc galaxies; a moderate correlation has been found in KMs. Our results for MMs and disc masers do not agree with previous studies, and they instead add another questionable piece to the AGN-maser relation.

Firstly, the absence of direct AGN-maser correlation may be caused by the complexity and variety of the conditions in which masers are triggered, which are still ambiguous. Most of the 200 known maser galaxies have an uncertain classification; while disc masers are most likely to be driven by AGN power, maser emissions in merger, jet, or heavy star formation rate galaxies are not dynamically related to AGN. Accordingly, we tried to check if merger (Mrk 1066, NGC 2146, NGC 3256, NGC 5256, Mrk 266, and NGC 6240), jet (NGC 4261, NGC 2824, NGC 1052, NGC 1194, Mrk 348, NGC 2639, NGC 1068, and NGC 6300), and outflow (Circinus, NGC 1068, Mrk 348, Mrk 1210, and NGC 1052) galaxies have any biasing effect on the results; excluding one or more of these galaxies does not change the results. Nevertheless, a precise classification of masers is not yet available, and it seems that confident classification of maser types (i.e., disc, star formation, outflow, or jet-powered masers), when available, will help figure out the contradictory results we get from our analysis.

On the other hand, maser flux density exposes up to three order of magnitude variability over timescales of months or even days (e.g., Braatz et al., 2003; Herrnstein et al., 2005; Castangia et al., 2007; Zhu et al., 2011; Pesce et al., 2015). This variability may have scattered our results and thus led to a correlation absence in MMs and disc masers.

Secondly, calculating $M_{BH}$ in maser galaxies using σ-$M_{BH}$ has a presumption that the CND's self-gravity is negligible, which may cause deviations in the results (Huré et al., 2011). However, Kuo et al. (2018) applied the model suggested by the last mentioned paper to 3-dimensions instead of 2-dimensions projection, and they found that for the three maser galaxies they studied, neglecting disc mass has a negligible effect. If this result is not similarly applied to all galaxies in our sample, then neglecting the disc's self-gravity might have caused a deviation in $M_{BH}$ values, future work of MCP's team will help clear out this doubtfulness.

Additionally, the σ-$M_{BH}$ relation has its challenges. Some theoretical works have proved that galaxy morphology and environment may change the way a galaxy follows the $M_{BH}$–σ relation (e.g., Hu, 2008; Wyithe, 2006). This can be unsafe for our analysis, especially if we take into account that water megamaser galaxies are found mostly in early to mid-type spiral galaxies and with $M_{BH}$ of around $10^7 M_\odot$ (Greene et al., 2010), both of which have significant deviations from the σ-$M_{BH}$ relation, as illustrated in Fig. 8 of Greene et al. (2010), and Fig. 3 of Greene et al. (2016). In these figures, maser galaxies lie below the σ–$M_{BH}$ line, with the highest offset for $M_{BH} < 10^7 M_\odot$ maser galaxies, which is the case for eight MMs and two disc maser galaxies of our sample, thus this may have significantly affected our results regarding the MMs sample. For the disc sample, fortunately, very precise $M_{BH}$-s measurements have been done using the Very Long-Baseline Interferometry (VLBI) techniques; an updated list of such measurements is reported in Table 5 of Kuo et al. (2020). We used the 21 disc maser galaxies' $M_{BH}$s in that table to examine if errors from the σ–$M_{BH}$ method are the reason behind the correlation absence. Surprisingly, our analyses show no correlation at all (r=-0.11, p = 0.65, and ρ = 0.36, p = 0.10). If the theory of Neufeld et al. (1994) is true, then we have three possible explanations for this. One possibility is that the current maser disc galaxy sample somehow has a complex structure of masing gas and thus does not follow the masing mechanisms described by Neufeld et al. (1994). Another possible explanation is related to the geometric alignment of the masing gas disc. Most maser flux is produced in the mid-plane where dense gas is concentrated; the disc has to be warped, so the mid-plane is not shielded from direct X-ray and thus gets heated enough (T > 400 K) to mase. The warping degree affects the heating efficiency, which may veil the X-ray-maser relation. Finally, assuming direct $L_{AGN}L_{X-ray}$ relation may not be met in all AGNs, some AGN flux can increase without increasing the X-ray output Kuo et al. (2018).

Notwithstanding the complications of using the $M_{BH}$–σrelation, it is still the most suitable way for statistical studies. Unfortunately, using other methods, such as reverberation mapping or the very accurate megamaser-based $M_{BH}$ measurements, is not currently available for enough samples.

## 7. Conclusion

Parametric and non-parametric correlation and linear regression methods were used to study the relation between maser luminosity and central black hole mass in 73 maser galaxies. Similar to previous studies, our results show a moderate and low significant correlation in KMs ($\rho_{partial}$=0.60, p = 0.012). However, for the MMs sample, we found no correlation (r = 0.17, p = 0.20). The correlation in the disc galaxies sample, expected to be the highest, is rather very weak and of low significance ($\rho$ = 0.41 and p = 0.03).

The regression line fits for these correlations have been calculated; non of the three samples (KMs, MMs, or disc masers) do we find the theoretical expected quadratic relation.

Trying to exclude one or more of merger, jet, and outflow maser galaxies -as they may bias the calculation- does not change the results. However, identifying maser types is not clear enough at the moment. It seems that what is defined as "clean disc masers" by Gao et al. (2016) still have some lurking merger, jet, and/or outflow masers.

Future samples with confirmed classification may allow explicit information for sample selection, which will clear up ambiguity. Moreover, the variability of maser fluxes may be the source of significant errors. Another possibility for the absence of tight correlation in MMs and disc masers could come from the deviation of $M_{BH}$ of these masers from the M-σrelation (see the results of Greene et al., 2016). Under other conditions, the fact that maser is not powered directly by AGN can not be excluded, and this suggestion is supported by the results of a couple of papers that studied $L_{H_2O}$-$L_{X-ray}$ relation with no positive results.

## Declaration of Competing Interest

The authors declare that they have no known competing financial interests or personal relationships that could have appeared to influence the work reported in this paper.

## Acknowledgments

The authors would like to thank the anonymous referee for thier valuable comments and insights that are brought to the paper.

This work has made use of the NASA/IPAC Extragalactic Database (NED), which is operated by the Jet Propulsion Laboratory, California Institute of Technology, under contract with the National Aeronautics and Space Administration. This work is supported by Bogazici University BAP under project numbers 13760 and 122F305, and ENE would like to thank them for their support.

## Appendix A. The 73 maser galaxies data sample

The data set used in this paper is presented in Table A.1. Columns shown in the Table are: (1). Name of the galaxy. (2). Type of the galaxy (see Section 3 for further details). (3). Redshift z, taken from Preferred redshift from NASA/IPAC Extragalactic Database (NED) (https://ned.ipac.caltech.edu/). (4). Isotropic $H_2O$ maser luminosity logarithm, $\log(L_{H_2O}/L_\odot)$. (5). $\log(\sigma)$. (6). $M_{BH}$ logarithm, $\log(M_{BH}/M_\odot)$. (7). References for $\sigma$, and $L_{H_2O}$, respectively.

Table A.1
The 73 maser galaxies sample.

| Galaxy name | Type | z | $\log(L_{H_2O})$ | $\log(\sigma)$ | $\log(M_{BH})$ | References |
|---|---|---|---|---|---|---|
| NGC 235A | MM | 0.0222 | 1.613 | 2.290 | 8.086 | 1, 8 |
| NGC 253 | KM | 0.0009 | −0.70 | 1.98 | 6.84 | 1, 8 |
| Mrk 348 | MM | 0.0153 | 2.60 | 2.02 | 7.00 | 1, 8 |
| ESO 013-G012 | MM | 0.0168 | 2.70 | 2.21 | 7.76 | 4, 8 |
| Mrk 1 (NGC 449) | MM | 0.0160 | 1.70 | 1.92 | 6.60 | 1, 8 |
| NGC 520 | KM | 0.0071 | 0.00 | 2.18 | 7.64 | 1, 8 |
| NGC 591(Mrk 1127) | MM, disc | 0.0153 | 1.40 | 1.96 | 6.76 | 1, 8 |
| NGC 613 | MM | 0.0050 | 1.30 | 2.09 | 7.28 | 1, 8 |
| Mrk 1029 | MM, disc | 0.0304 | 2.83 | 2.12 | 7.40 | 2, 8 |
| NGC 1052 | MM | 0.0049 | 2.10 | 2.32 | 8.21 | 1, 8 |
| NGC 1068 | MM, disc | 0.0038 | 2.20 | 2.21 | 7.76 | 1, 8 |

Table A.1 (*continued*)

| Galaxy name | Type | z | log($L_{H_2O}$) | log($\sigma$) | log($M_{BH}$) | References |
|---|---|---|---|---|---|---|
| NGC 1106 | KM | 0.0145 | 0.903 | 2.165 | 7.583 | 1, 8 |
| Mrk 1066 | MM | 0.0121 | 1.505 | 2.021 | 7.005 | 4, 8 |
| NGC 1194 | MM, disc | 0.0136 | 2.049 | 2.170 | 7.603 | 2, 8 |
| NGC 1320 | MM, disc | 0.0090 | 1.633 | 2.041 | 7.086 | 1, 8 |
| NGC 1386 | MM, disc | 0.0029 | 2.1 | 2.079 | 7.238 | 4, 8 |
| IC 342 | KM | 0.0001 | $-2$ | 1.869 | 6.394 | 1, 8 |
| J043703.67 + 245606.8 | MM, disc | 0.0162 | 2.25 | 2.04 | 7.08 | 2, 8 |
| Mrk 3 (UGC 3426) | MM | 0.0135 | 1.00 | 2.39 | 8.49 | 1, 8 |
| NGC 2146 | KM | 0.0030 | 0.00 | 2.1 | 7.32 | 1, 8 |
| NGC 2273 | MM, disc | 0.0062 | 1.51 | 2.15 | 7.52 | 1, 8 |
| ESO 558-G009 | MM, disc | 0.0259 | 2.88 | 2.23 | 7.84 | 2, 8 |
| UGC 3789 | MM, disc | 0.0107 | 2.57 | 2.03 | 7.04 | 2, 8 |
| Mrk 78 | MM, disc | 0.0380 | 1.51 | 2.28 | 8.05 | 1, 8 |
| Mrk 1210 | MM, disc | 0.0136 | 1.90 | 2.06 | 7.16 | 4, 8 |
| NGC 2639 | MM | 0.0107 | 1.40 | 2.24 | 7.88 | 1, 8 |
| NGC 2781 | MM | 0.0068 | 1.146 | 2.068 | 7.193 | 1, 8 |
| NGC 2782 | MM | 0.0085 | 1.11 | 2.26 | 7.97 | 1, 8 |
| NGC 2824 | MM, disc | 0.0092 | 2.70 | 2.1 | 7.32 | 1, 8 |
| NGC 2979 | MM, disc | 0.0090 | 2.10 | 2.05 | 7.12 | 4, 8 |
| M 82 (NGC 3034) | KM | 0.0007 | 0.00 | 2.1 | 7.32 | 1, 8 |
| NGC 3081 | MM | 0.0080 | 1.23 | 2.076 | 7.225 | 1, 8 |
| NGC 3079 | MM, disc | 0.0038 | 2.70 | 2.24 | 7.88 | 1, 8 |
| IC 2560 | MM, disc | 0.0098 | 2.00 | 2.13 | 7.44 | 2, 8 |
| NGC 3256 | MM | 0.0094 | 1 | 2.037 | 7.069 | 1, 8 |
| Mrk 34 | MM, disc | 0.0510 | 3.00 | 2.26 | 7.97 | 4, 8 |
| NGC 3359 | KM | 0.0034 | $-0.15$ | 1.74 | 5.87 | 1, 8 |
| NGC 3393 | MM, disc | 0.0126 | 2.51 | 2.29 | 8.09 | 1, 8 |
| UGC 6093 | MM, disc | 0.0360 | 2.89 | 2.19 | 7.68 | 2, 8 |
| NGC 3556 | KM | 0.0023 | 0.00 | 1.9 | 6.52 | 1, 8 |
| NGC 3735 | MM | 0.0090 | 1.30 | 2.15 | 7.52 | 1, 8 |
| NGC3783 | MM | 0.0098 | 1.34 | 2.05 | 7.12 | 1, 8 |
| NGC 4051 | KM | 0.0022 | 0.30 | 1.96 | 6.76 | 1, 8 |
| NGC4151 | KM | 0.0033 | $-0.22$ | 1.96 | 6.76 | 1, 8 |
| NGC 4194 | MM | 0.0082 | 1.079 | 1.996 | 6.904 | 1, 8 |
| NGC 4214 | KM | 0.0010 | $-1.52$ | 1.71 | 5.75 | 5, 8 |
| NGC4253 | KM | 0.0127 | 0.95 | 1.97 | 6.80 | 6, 8 |
| NGC 4258 | MM, disc | 0.0015 | 1.90 | 2.12 | 7.40 | 1, 8 |
| NGC 4261 | MM | 0.0073 | 1.60 | 2.47 | 8.81 | 1, 9 |
| NGC 4293 | KM | 0.0030 | 0.00 | 2.07 | 7.20 | 1, 8 |
| NGC 4388 | MM, disc | 0.0084 | 1.08 | 2 | 6.92 | 1, 8 |
| NGC 4527 | KM | 0.0058 | 0.60 | 2.13 | 7.44 | 1, 8 |
| ESO 269-G012 | MM, disc | 0.0169 | 3.00 | 2.21 | 7.76 | 3, 8 |
| NGC 4922 | MM | 0.0234 | 2.30 | 2.33 | 8.25 | 1, 8 |
| NGC 4945 | MM | 0.0018 | 1.70 | 2.07 | 7.20 | 1, 8 |
| NGC 4968 | MM | 0.0100 | 1.724 | 2.013 | 6.972 | 1, 8 |
| NGC 5077 | MM | 0.0090 | 1.85 | 2.4 | 8.53 | 1, 9 |
| NGC 5128 | KM | 0.0018 | 0.00 | 2.01 | 6.96 | 1, 9 |
| M 51 (NGC 5194) | KM | 0.0013 | $-0.22$ | 1.94 | 6.68 | 1, 8 |
| NGC 5256 (Mrk 266) | MM | 0.0274 | 1.51 | 2 | 6.92 | 7, 8 |
| NGC 5347 | MM | 0.0080 | 1.51 | 1.97 | 6.80 | 1, 8 |
| NGC 5495 | MM, disc | 0.0225 | 2.78 | 2.22 | 7.80 | 2, 8 |
| Circinus | MM, disc | 0.0014 | 1.30 | 2.17 | 7.60 | 1, 8 |
| NGC 5506 | MM | 0.0059 | 1.70 | 2.24 | 7.88 | 4, 8 |
| NGC 5643 | MM | 0.0040 | 1.40 | 1.97 | 6.80 | 4, 8 |
| NGC 5728 | MM, disc | 0.0095 | 1.90 | 2.29 | 8.09 | 1, 8 |
| NGC 5765b | MM, disc | 0.0275 | 3.36 | 2.21 | 7.76 | 2, 9 |
| NGC 6240 | MM | 0.0243 | 1.60 | 2.53 | 9.05 | 1, 8 |
| NGC 6264 | MM, disc | 0.0337 | 3.32 | 2.2 | 7.72 | 2, 8 |
| NGC 6323 | MM, disc | 0.0259 | 2.70 | 2.2 | 7.72 | 2, 8 |
| NGC 6300 | KM | 0.0037 | 0.48 | 2.01 | 6.96 | 4, 8 |
| ESO 103-G35 | MM | 0.0132 | 2.60 | 2.06 | 7.16 | 4, 8 |
| NGC 7479 | MM | 0.0080 | 1.28 | 2.18 | 7.64 | 1, 8 |

**References.** $\sigma$data: (1). http://leda.univ-lyon1.fr/a001/ (2). Sahu et al. (2019). (3). Fernandes et al. (2004). (4). Su et al. (2008). (5). Ho et al. (2009). (6). Woo et al. (2015). (7). Dasyra et al. (2011). $H_2O$ data: (8). The MCP web page "https://safe.nrao.edu/wiki/bin/view/Main/PublicWaterMaserList". (9). Kuo et al. (2018).